**LoopX: Visualizing and understanding the origins of dynamic model behavior.**
By William Schoenberg (University of Bergen, Norway)

**Abstract:**
It is a fundamental precept of System Dynamics that structure leads to behavior. Relating the two is a roadblock for using feedback models because substantial experimentation or the application of specialized analytic techniques that are difficult to employ is required. LoopX builds understanding of structure as it determines behavior by rendering and highlighting structure responsible for behavior as it unfolds. LoopX builds on the Loops that Matter (Schoenberg et. al, 2019) approach to analyzing loop dominance by presenting its results in an easy to use, interactive, software tool. This is a significant step forward in the challenges of automatically visualizing model behavior and linking it to generative structures identified in Sterman (2000). LoopX machine generates high quality CLDs from model equations at different levels of detail based on the dynamic importance of links and variables, in addition to animating them based on their importance to the origins of model behavior.

**Introduction**
This paper presents a new and highly usable solution to three important challenges identified in the final chapter of *Business Dynamics* (Sterman 2000): It addresses "Automated identification of dominant loops and feedback structure", calculating and displaying the evolution of loop dominance as a model simulates; It improves on "Visualization of model behavior", using animation to coordinate the display of structural dominance evolution with the behavior over time of model variables; It addresses "Linking behavior to generative structure", using animation of automatically aggregated diagrams that connect the loop dominance analysis with the model structure via connectors and flows that change size and colors over the course of a model simulation.

Simple systems are usually easy to analyze with intuition and trial-and-error, but with larger systems that are characterized by high feedback loop complexity, the risk of incorrect explanation rises (Oliva, 2016). It is this threat of failure which makes these three challenges posed by Sterman (2000) so relevant. Currently the domain of objective feedback loop dominance analysis is limited to a relatively select few practitioners with a high degree of expertise and training. The lack of tools for parsing and developing insight in large causal models often acts as the limit on the utility of large models to general audiences (Schoenenberger et. al, 2017). The incidence of these problems with presenting models with the intent to develop understanding is not a new occurrence, a cursory literature returns a 1976 paper (republished in 1986) which refers to problems in methods for simplifying the presentation of model structure via casual loop diagrams developed even earlier than that (Richardson, 1986). Cleary, any solution to Sterman's three challenges must help to reduce the barriers to entry for model understanding and analysis, expanding our depth of understanding of the models which are at the heart of our field via improved communication of complexity and its origins.

The foundation of this work is the Loops that Matter (LTM) technique for determining loop dominance (Schoenberg et. al, 2020). Building on the LTM method, the solution to these three challenges employs the use of Causal Loop Diagrams (CLDs) as well as Stock and Flow Diagrams (SFDs) as a vehicle for representing system structure to model consumers.

This paper presents LoopX which is a tool that is capable of reading in and analyzing an XMILE model. The tool allows the model to be simulated, and also analyzed by LTM generating a full complement of link and loop scores describing the origins of model behavior from a loop dominance perspective. LoopX is capable of rendering the model as a stock and flow diagram based on the layout decisions made by the model author. LoopX also machine generates high quality CLDs from the network of model equations at user specified levels of complexity presenting a minimum number of variables and links which are deemed necessary (by the LTM analysis) to understand the dynamics of the shifting loop dominance at the requested cognitive complexity level. All diagrams, machine or human generated, are animated portraying dominance information via flows and connectors which change colors and size in real time as the model simulates. All loops are identifiable directly within the context of all of the aforementioned diagrams.

**Problem Statements**
LoopX required the development of solutions to the following three main problems:

1. How can high quality CLDs be machine generated from the network of model interconnections?
2. How can models be aggregated and simplified without losing information important to model understanding while retaining relative simplicity?
3. How can the results of an LTM loop dominance analysis be easily visualized and communicated?

**Literature Review**
This review combines literature from the graph theory and system dynamics fields to provide the reader with the requisite knowledge for understanding the current state of the art as it applies to each of the three problem statements. This helps to place the development of LoopX into context among the existing technologies.

**Techniques for machine generation of network graphs, the basis on which CLDs are formed**
Most important to the automated generation of high quality CLDs is the force directed layout algorithm. A force directed layout algorithm solves the problem of the placement of nodes in 2D space, such that symmetry is generated, and edge length is approximately equal, by running a physics simulation of weights connected by springs and minimizing the total energy of the system. The first force directed layout technique used steel rings to represent each node and then connected those rings using logarithmic springs (Eades, 1984). In this version of the algorithm, attractive forces were only calculated between neighbors, and repulsive forces were calculated between all node pairs (Eades, 1984). This process ensured that neighbors were always close by but limited the scope of the N-squared problem.

The next evolution in the force directed algorithm was to introduce the concept of an ideal distance between every node pair based on the shortest path between each node pair, and to use Hooke's Law, meaning real world realistic linear springs (Kamada and Kawai, 1989). The Kamada Kawaii approach solved partial differential equations based on Hooke's Law to optimize layout applying all forces between all node pairs in an iterative fashion (Kamada and Kawai, 1989).  A gradient descent optimization process used to terminate the simulation when a global minimum in the energy state of all the springs was found (Kamada and Kawai, 1989).

Development of Graphviz, an open source toolkit for solving these graph generation problems took place in parallel to these developments at Bell Labs.  Graphviz contains many different automated layout mechanisms, but the mechanism most relevant to CLD generation is called neato, which is based on the work of Eades, Kamada and Kawai among others.  The layout algorithm used in neato that we are interested in, is derived from the Kamada Kawai algorithm. It assumes there is a linear spring between every pair of nodes, each with an ideal length (Gansner, 2014). The ideal distance between each node pair is the result of a function computed for each pair; the ideal length function we are interested in uses the shortest path between the two nodes to determine the ideal distance between these nodes, although many other choices are offered.  Neato is able to turn a static text file with a description of the graph as lists of nodes and edges into a 2D diagram quickly (North, 2004).

Neato performs the following series of high-level steps in its operation.  First neato parses its input file which specifies the list of nodes and all edges which connect them, as well as any arguments which are used to control the layout process.  Second it constructs and simulates the physics simulation to lay out the nodes in 2D space.  Neato allows the user to specify the starting location of each node, and since the layout process relies on a gradient descent optimization this allows the user to potentially identify starting locations for the nodes which produce higher quality diagrams then others.  At this point neato executes any node overlap removal operations.  Finally, once all nodes are laid out, neato draws the edges connecting the nodes according the options specified in the input file.

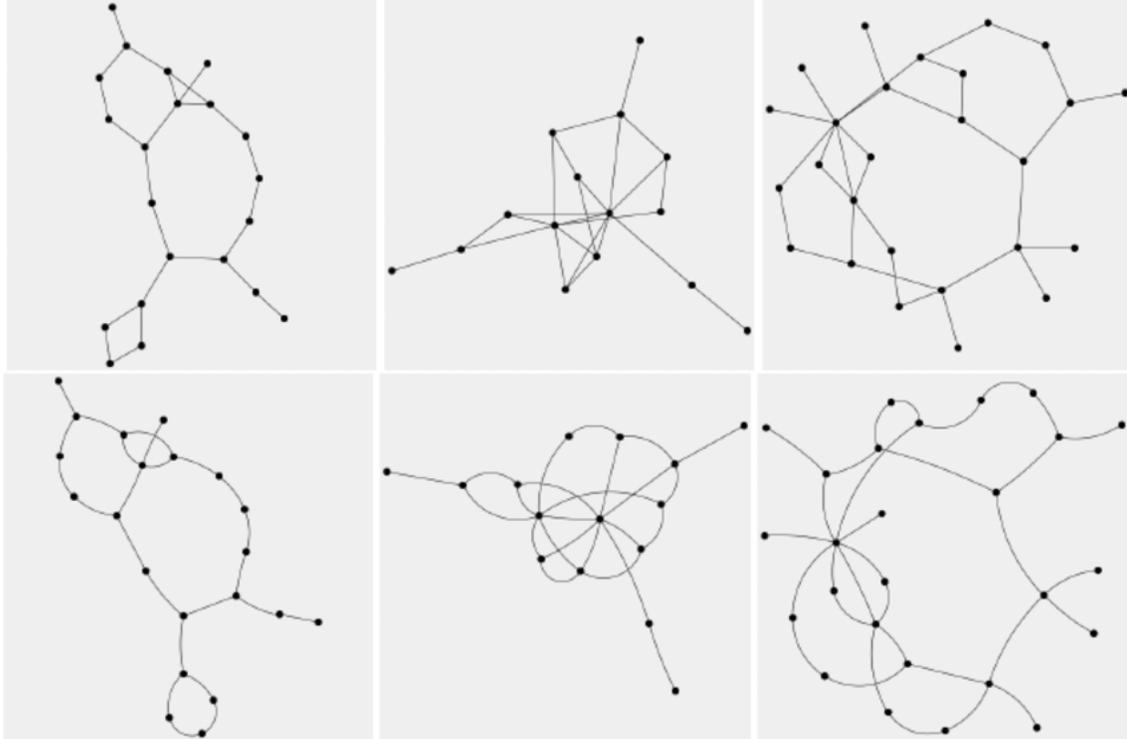

*Figure 1: Example of standard straight-line and Lombardi-style force directed graphs copied from Chernobelskiy et al 2011.*

Neato, like all force directed graph algorithms, produces, by default, edges that are straight. There are disadvantages though for using straight edges especially as it relates to user understanding of the generated graph diagrams (Xu et al. 2012). Graphs of the type produced by force directed algorithms with curved edges are generally called near-Lombardi or Lombardi-style diagrams, and CLDs are a form of Lombardi-style diagram. Figure 1, presents examples of straight edge force directed graphs and their Lombardi style complements. When using a curved edge, Lombardi-style diagram, there are significant user performance improvements on graph related tasks such as determining the shortest path, node degree calculations and common neighbor determinations (Xu et. al., 2012). The problem with algorithmic Lombardi-style diagram generation and force directed graphs in general, is the lack of the concept of directed edges and especially directed cycles (links and loops as SD conceives of them). Lombardi-style diagrams tend to produce circular shapes that are not loops. To meet the needs of the SD community, and to be able to apply force directed graph algorithms like neato to generating CLDs, an algorithm for determining how to curve each edge in a way which emphasizes the loops needed to be developed.

**Techniques for model aggregation and simplification**
As covered in the introduction, the problems of simplifying dynamic complexity for wider consumption has been studied since the formative years of the field. Early research discusses how CLDs alone do not give an accurate enough picture of model structure so that behavior modes can be predicted and understood (Richardson, 1986). Richardson (1986) argues for caution when using CLDs to aggregate and simplify model diagrams, and that information is often lost in that process.

The most famous examples of feedback simplification techniques are the independent loop set and its refinement, the shortest independent loop set (ILS and SILS) respectively (Kampman, 2012) and (Oliva, 2004). These graph theory techniques for the partitioning of the cycles (feedback loops), implicit in the network of model structure, arose due to the complexity faced when performing and analyzing the results of the eigenvalue elasticity method of loop dominance analysis (EEA) or when trying to find high leverage points for policy intervention. In anything but small models both authors were faced with a relatively large list (compared to the number of variables in the model) of feedback loops which were all tightly interrelated. The SILS concept pairs down the number of feedback loops to the set of geodetic (shortest) loops which are necessary to fully describe the feedback loop complexity of the model. This reduces the number of loops present in a fully accurate CLD of the entire network of model structure, focusing user attention on the loops which are most easily influenced by policy.

Built on the ILS and SILS, (Schoenenberger et. al, 2015 and 2017) present the use of variety filters derived from interpretative model partitioning, structural model partitioning and the ADAS method (algorithmic detection of archetypal structures), to communicate intuition from large models. Their audience are those who would normally be overwhelmed by the size and complexity of the models being studied. This work also builds upon earlier studies of model simplification done by Eberlein (1989), which uses linearization, and on Saysel and Barlas' (2006) aggregation method. The variety filters technique presents the user with structural clusters of model variables based on state of the art statistical and graph theory techniques as a way of visualizing and understanding nearness and hierarchy. With interpretative clustering, model complexity is filtered via studies of the relationships between pairs of model sectors. Using ADAS which is applied to the above generated clusters, users select a stock of interest as well as an archetypal structure to find, and the algorithm returns the feedback loops which contain the variable in the system archetype specified. This significantly reduces the number of feedback loops to be studied by the end user pairing down the complexity of the model.

The Forio Model Explorer feature of Forio Simulate is an example of a simplistic Kamada Kawai style force directed rendering of model structure which was later evolved into supporting a secondary hierarchical layout engine with rudimentary aggregation steps taken to either only show two degrees of distance from a variable of interest or all of the links between two variables of interest with a filter based on path length. The Forio Model Explorer was studied and was compared to traditional hand drawn CLDs in an attempt to measure the effectiveness of the automated diagramming and aggregation techniques (Schoenberg, 2009). All tests were inconclusive, showing no reported differences in learning outcomes, but diagrams generated were of significantly less quality, lacked any of the positive attributes of Lombardi-style diagrams and were not focused on feedback loop behavior. This work appears to be the most recent previous attempt at using aggregation and force directed graphs to solve the challenges laid out by Sterman (2000).

**The Loops That Matter method**
The LTM method (Schoenberg, et. al, 2020) performs a formal assessment of dominant structure and behavior as categorized by Duggan and Oliva (2013). The LTM method is built around the observation of how modelers perform the art of model analysis to understand the origins of behavior. LTM interacts directly with the full network of model equations, walking the causal pathways between all variables in the model, calculating in time with the simulation, metrics that measure the contribution (ex: force, strength) and polarity of each link in the network of model equations. The LTM approach produces metrics which interpret the origins of behavior for the entire model[1] rather than just the behavior of a single state variable.

The first metric introduced by the LTM method is the link score. The link score is a measure of the contribution and polarity of any link in a model from an independent to dependent variable regardless of whether or not the link contains an integration process. The link score concept tracks the concept of the link gain, and when multiplied through pathways up until but not including the stock, is the same as Richardson's (1995) concept of the dominant polarity. The link score is capable of being calculated for every link in the mode, including those which contain an integration process. The link score is computed once per each time interval in the model and is computed for each link in the model. There are two methods for calculating the link score depending upon if the link contains an integration process or not. Schoenberg et. al. (2020) demonstrates that the two methods produce exactly the same measure and therefore can both be referred to as the link score.

*Equation 1* is the definition of the link score of a link that does not contain integration assuming there are two inputs ($x$ and $y$) to the dependent variable z characterized by the equation $z = f(x, y)$. The link score for the link x → z when written in a discontinuous form based upon the implementation of the calculation is (See Schoenberg et. al, 2020 for continuous analytical form):

*Equation 1: The discontinuous form for the link score equation which matches how the implementation of the calculation works moving in time with the dt of the model.*

$$LS(x \rightarrow z) = \begin{cases} \left( \left| \frac{\Delta_x z}{\Delta z} \right| \cdot sign\left( \frac{\Delta_x z}{\Delta x} \right) \right), \\ 0, \quad \Delta z = 0 \text{ or } \Delta x = 0 \end{cases}$$

In *Equation 1* Δz is the change in z from the previous time to the current time. Δx is the change in $x$ over that same time step. $\Delta_x z$ is the change in $z$ with respect to $x$. From a computational perspective $\Delta_x z$ which is called the partial change in $z$ with respect to $x$, is the amount $z$ would have changed, conditionally, if $x$ had changed the amount it did, but $y$ had not changed. The first major term in *Equation 1* represents the magnitude of the link score, the second is the link score polarity.

---

[1] For cases where each stock in the model is able to either directly or indirectly impact each other stock in the model. For models where this assumption does not hold true LTM informs on the origins of behavior in each giant connected component of the model where this assumption holds true.

*Equation 2: Link score for all links from derivatives (flows) to state variables (stocks) (both inflows and outflows are covered). The simple one inflow and one outflow case is presented and is easily generalized.*

$$Inflow: LS(i \rightarrow s) = \left(\left|\frac{i}{i-o}\right| * 1\right) \quad Outflow: LS(o \rightarrow s) = \left(\left|\frac{o}{i-o}\right| * -1\right)$$

For links which contain an integration process the equation for determining the contribution and polarity of the link from an inflow (*i*) or outflow (*o*) to a state variable (*s*) is shown in *Equation 2*. This allows the LTM method to measure the link score which is the contribution and polarity of each individual flow to the value of the stock.

The second key metric produced by the LTM method is the loop score as shown in *Equation 3*. The loop score tracks the concept of the loop gain and is the result of the multiplication of all link scores for all links in a loop. This is a demonstrably unique measure which bears some rough similarity to the Loop Impact metric of Hayward and Boswell (2014) but is unique because it is capable of including the links which contain integration processes allowing a single value to be assigned as the contribution of a loop.

*Equation 3: Definition of loop score, for the loop x which contains n links for each source variable S to the target variables T.*

$$Loop\ Score(L_x) = \left(LS(s_1 \rightarrow t_1) \cdot LS(s_2 \rightarrow t_2) \ldots \cdot LS(s_n \rightarrow t_n)\right)$$

The loop score is a dimensionless value which samples the effort a loop is expending to change the behavior of the stocks it connects at each calculation interval of the model. As the link score can be thought of as the force of an independent variable pushing on the result of a dependent variable, the loop score can be thought of as the force of one feedback loop pushing on the behavior of all the stocks (and therefore all variables) it connects.

The third and final key metric produced by the LTM method is the relative loop score (*Equation 4*) which compares the contribution of feedback loops to determine which are dominant at any point in time. The relative loop score requires no independence across the loops it compares and ideally uses the exhaustive set of feedback loops as the basis for comparison.

*Equation 4: Definition of the relative loop score for the loop x normalized over all loops n analyzed in the chosen loop set.*

$$Loop\ Score_{L_x} = \left(\frac{Loop\ Score(L_x)}{\sum_{y=0}^{n}|Loop\ Score(L_y)|}\right)$$

The sign of a relative loop score represents the polarity of the feedback loop. The relative loop score is a normalized measure taking on a value between -1 and 1. It reports the polarity and instantaneous fractional contribution of a feedback loop to the change in value of all stocks in the feedback loop set it is a member of. By comparing loop scores, it can easily be determined which loops are dominant, i.e. contribute the most (over 50%) to the behavior of all stocks in the feedback loop set under study. This normalization is critical to maintaining scores that are easy to work with.

**Overview of the LoopX software**

Figure 2 demonstrates a high-level organization of the subprocesses inside of the LoopX software. The LoopX is a web-based software tool capable of reading in an XMILE model file which must contain a stock and flow diagram. The LoopX tool is capable of simulating a limited set of XMILE models and performing a loop dominance analysis on those models using LTM. The LoopX tool is also capable of rendering models as either animated SFDs or CLDs based upon the loop dominance analysis.

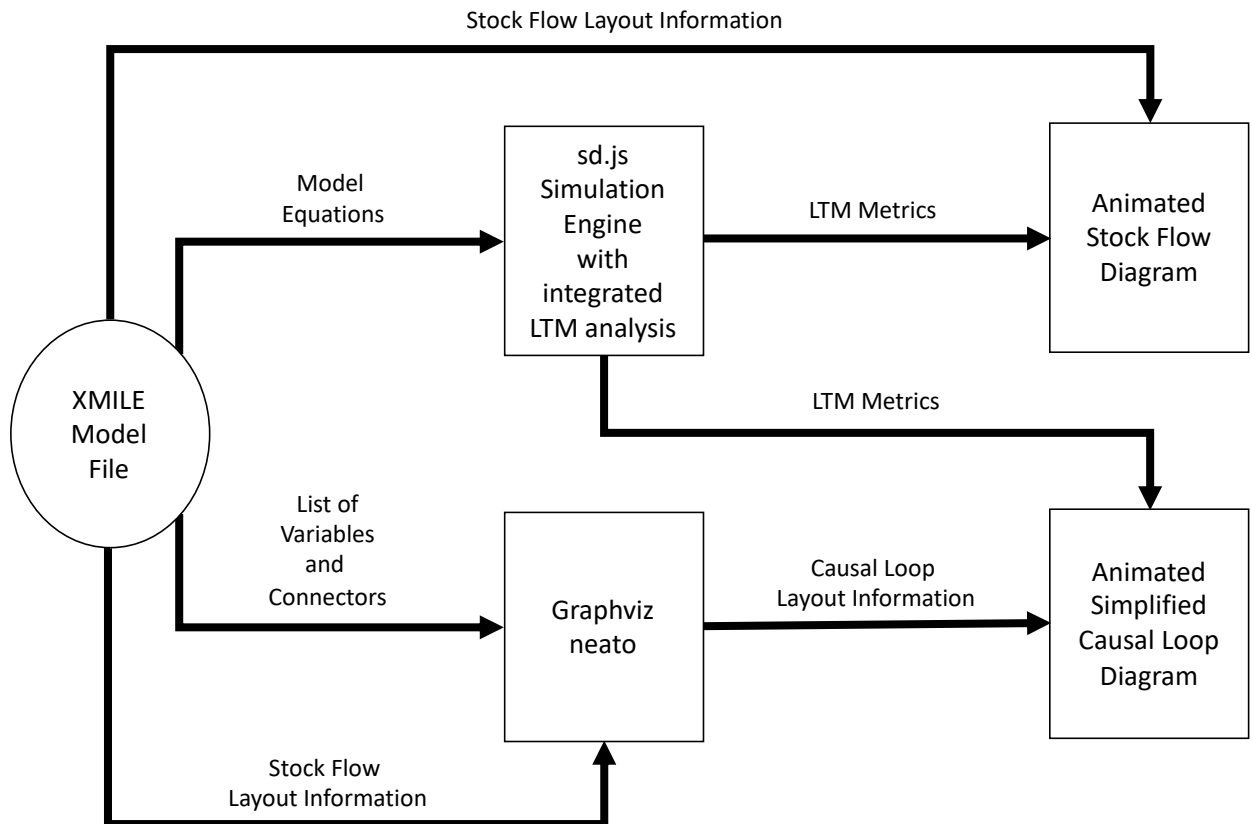

*Figure 2: Schematic showing the relationship of the various processes in the LoopX software. Each arrow is labeled with the information that passed between processes (the squares).*

LoopX starts with a user provided XMILE model file. Once the file is parsed the model equations are used by the sd.js simulation engine constructed by Powers (2019) which has been modified by Schoenberg et. al., (2020) to simulate the model while performing the LTM analysis. Each link in the model has its link score calculated for each dt. Each loop in the model has its loop score and relative loop score calculated for each dt in the model. These are the LTM metrics referred to in Figure 2. The model file is also used to generate a list of variables (nodes) and connectors (edges) and that information along with the initial position of the variables in the stock flow diagram is passed to neato which uses them to perform the automated CLD generation. Neato runs a Kamada Kawai force directed graph algorithm to produce the static CLD that is then animated by LoopX using the LTM metrics. The animated stock and flow diagram combines the static stock and flow diagram information from the XMILE file and the LTM metrics from the simulation engine.

**Solutions to the three major challenges**

1. *How can high quality CLDs be machine generated from the network of model interconnections?*

Neato is used to generate the static causal loop diagrams which are then later animated by LoopX to reveal the origins of model behavior based upon an LTM analysis. As identified in the literature review, the problem with using neato and force directed graph algorithms in general, is that they produce diagrams with all straight edges, or they produce diagrams with curved edges that do not emphasize the feedback loops. To develop machine generated CLDs which emphasize the loops, the development of LoopX required a solution to this problem. Figure 3 (as well as all of the other CLDs shown in this paper) demonstrate the solution to that problem.

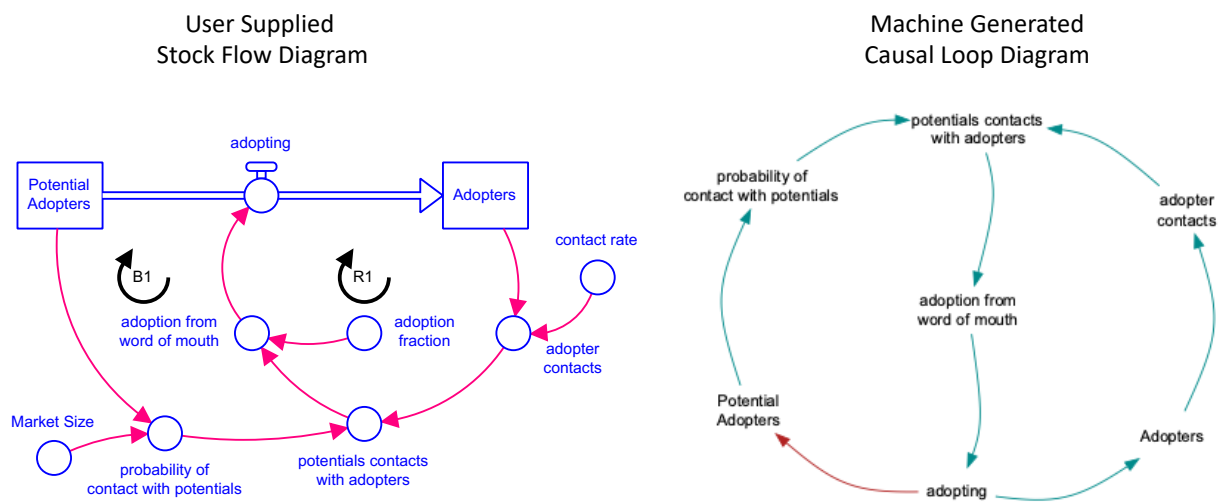

Figure 3: Example machine generated CLD from user supplied SFD of the Bass diffusion model (1969).

To make neato capable of drawing edges in a way that emphasizes the feedback loops, the edge drawing step of neato had to be modified. The solution to the edge curving problem is a simple algorithm whose implementation has been accepted into the publicly available version of neato and is invoked whenever the user specifies that they want their edges to be curved. The edge curving algorithm follows a simple heuristic derived from the observation of CLD diagram drawing by hand. The heuristic specifies that on an edge by edge basis, the center of the circle which forms the arc that the edge will follow, must be the average center of the nodes which form the shortest feedback loop with length greater than 2 that the edge is a member of. The two-node exception is handled separately within the neato codebase, and produces paired directed edges that do not over-emphasize the cycle, - producing elongated ellipse structures that cover an area relative to the number of nodes. This edge curving heuristic relies upon the attributes of force directed graphs which place nodes that are related closest together. This heuristic produces loops that look circular except for in degenerate cases where the force directed layout fails to produce good local clusters and the shortest feedback loops are relatively far flung in the 2D space.

To minimize the incidence of degenerate diagrams due to path dependency issues in neato use as the initial position of each node, the position of the variable it represents in the stock and flow diagram. This is a generally useful way to make sure that local clusters are being preserved which is important for keeping short loops near each other in the machine generated diagram. This is based on the assumption of a good quality stock and flow diagram which has also kept the variables most closely related to each other near in physical space.

Finally, to raise the probability of generating high quality CLDs, - LoopX uses the following neato specific settings. First, LoopX instructs neato to use the Prism algorithm (Gansner & Hu, 2010) (a proximity graph-based algorithm) to prevent the overlap of variable names in the CLD by setting the 'overlap' attribute of the graph to 'prism'. This instructs neato to remove overlapping variable names in a way which minimizes the disturbance to the layout created by the physics simulation. Second, LoopX instructs neato to use the KK mode during the diagram layout step so that neato uses a variant of the gradient descent process originally proposed by Kamada and Kawai (1989), for solving the optimization problem during the node placement step. This is done by setting the graph attribute 'mode' to 'KK'. Third, LoopX instructs neato to set the ideal edge length based on the 'shortpath' model (default option) for computing its distance matrix, i.e. using the shortest path between two nodes as the ideal length of the spring between each node pair. This is done by setting the graph attribute 'model' to 'shortpath'. Finally, LoopX invokes the edge curving algorithm for emphasizing feedback loops by setting the graph option 'splines' to 'curved', - otherwise a straight-edged diagram would be produced.

2. *How can models be aggregated and simplified without losing information important to model understanding while retaining relative simplicity?*

Based on the LTM loop dominance analysis, LoopX introduces two new parameters which are used to specify how to simplify causal loop diagrams in a way which maximizes the explanatory nature of the diagrams while removing variables and feedback complexity. The parameters are used to filter the full feedback complexity of the model, reducing the number of variables and therefore the links between variables in a CLD in such a way as to minimize the loss of descriptive power as measured by the relative loop score.

The first parameter is the 'link inclusion threshold' which filters variables from the simplified CLD by measuring the maximal variance in the contribution of links. To do that, a method must be derived to measure the variance in link contribution. The 'relative link variance' calculation is demonstrated in Equation 5 and measures the variance in the contribution of a link across the entire time period of the simulation. The relative link variance measures the change in the percentage contribution of an independent variable $x$, to a dependent variable $y$, across the entirety of a simulation run. The basis for the calculation is the relative link score which takes the link score metric from the LTM analysis and calculates a normalized value across all independent variables for a dependent variable.  The relative link score describes the contribution and polarity of a link as a percentage of the total contribution across all incoming links of a dependent variable.

*Equation 5: Relative Link Variance, calculated for the link x→y which measures the variance in the magnitude of the relative link score over the full simulation time.  The relative link score is the link score normalized across all determinants of a target variable.  In this case the relative link score is normalized over all determinants of y.*

$$\begin{aligned} Relative\ Link\ Variance(x \to y) \\ = \max\left(abs\big(Relative\ Link\ Score(x \to y)\big)\right) \\ - \min\left(abs\big(Relative\ Link\ Score(x \to y)\big)\right) \end{aligned}$$

The link inclusion threshold is used mainly to filter the number of auxiliary variables that appear in the simplified CLD rendered by LoopX.  The link inclusion threshold has a range of [0-1].  Only variables which are pointed to by at least one link with a relative link variance greater than or equal to this parameter, are included in the simplified CLD.  If a stock is included, LoopX automatically includes the flows to make clear to the end user that the stocks require flows to change.  The link inclusion threshold allows the user to specify from a loop dominance perspective which (mainly auxiliary) variables to remove from the simplified CLD.  The reason that typically only auxiliary variables are filtered by the link inclusion threshold is because flow-to-stock links typically have high relative link variance because there are large changes in link score as the model approaches and leaves equilibrium states.  Links with a high relative link variance are those which change their contribution the most over the course of the simulation run. Those links, therefore, tend to point to the sources of non-linearity in models and are, typically, variables that are important for understanding model behavior.  Links which do not change at all over the course of the simulation run, regardless of the specific level of contribution to their dependent variable, have a relative link variance of 0, and point to variables that are likely unimportant and are good candidates for elimination.  These variables tend to exist for the modeler to simplify equations.

The second parameter used to simplify CLDs is the 'loop inclusion threshold' which is mainly used to filter the number of stocks and associated flow variables that appear in the simplified CLD.  The loop inclusion threshold has a range of [0-1], representing the average magnitude of the relative loop score from LTM.  Only loops that have an average magnitude of the relative loop score greater than or equal to this parameter have their stocks and flows automatically included in the simplified CLD regardless of the link inclusion threshold filtering.  This allows the user to specify from a loop dominance perspective which stock and flows (as well as their associated direct feedback loops) do not need to be in the rendered graph, - facilitating the creation of a smaller simplified CLD.  The average magnitude of the relative loop score is measured over the entire time period of the simulation run, starting in the first instant where the loop becomes active.  This delayed averaging avoids penalizing loops during the initialization phase of model behavior.  The average magnitude of the relative loop score describes the percentage of the change in behavior of the stocks in the model which the loop is responsible for.  Loops with a low average magnitude of the relative loop score, for instance, 0.01 would only describe 1% of the total behavior of the model over the simulated time horizon and probably do not need to be included in a simplified diagram in order to understand the origins of the generated model behavior from a structural perspective.

The link and loop inclusion thresholds are applied to all links and loops in the model and produce a filtered list of variables to keep directly from the network of model equations.  From that, a corresponding list of links needs to be generated, - one that matches the reality of the true feedback loop connections in the fully disaggregated model but does not show all of the individual steps along the way.  In other words, the filtering process solves the aggregation problem from a variable perspective and now the links need to be generated such that they make sense at both this new level of aggregation and yet still represent the connections of a totally disaggregated model.

The process used to generate the links performs a depth first search for each possible link in the new limited set of variables generated by the filtering of the link and loop inclusion thresholds.  For each candidate simplified link the search traverses the full model, testing if it can walk the network of full model equations from the source of the candidate link to the target without passing through a variable already in the limited set specified by the parameter filters (or a variable already visited in the search).  If it can find a pathway in the full disaggregated model which properly represents the candidate link, the candidate link is kept.  All kept simplified links are both valid in the fully disaggregated model, because the search was able to find a causal chain which that link represents, and each simplified link only represents a single step in the aggregated model.

Next the application of this technique is studied in the context of Forrester's 1968 Market Growth model picture in Figure 4.  The Market Growth model with all macros expanded contains 23 feedback loops, 48 variables with 10 stocks.  The full CLD containing the full feedback complexity and network of equations involved in feedback loops is presented in Figure 5 and the first chosen level of simplification in Figure 6.

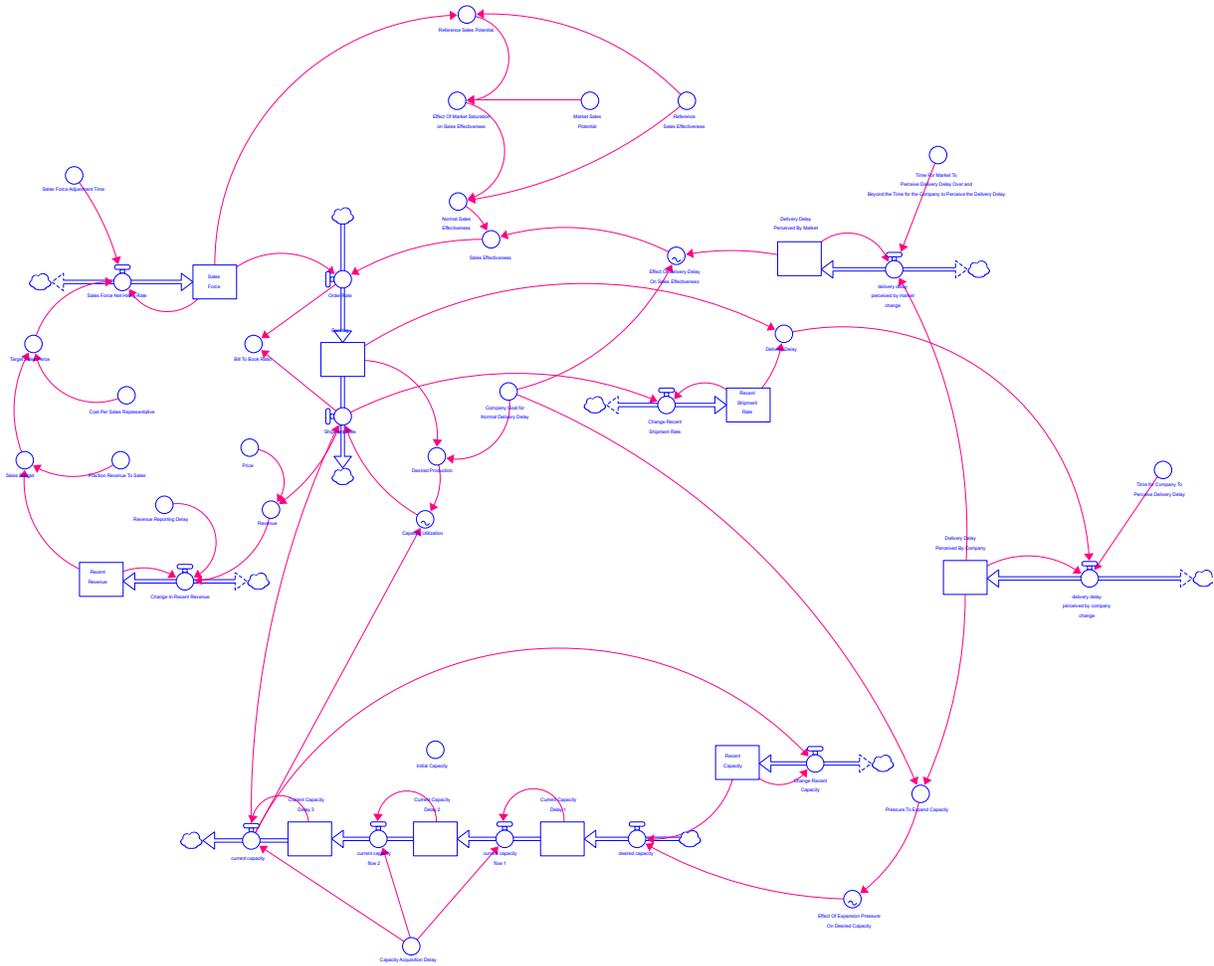

*Figure 4: Stock and flow diagram of Forrester's 1968 market growth model. Full model is included in the online supplemental materials*

*Figure 5: Autogenerated full CLD of Forrester's 1968 market growth model (link inclusion threshold 0%, loop inclusion threshold 0%.) Red links are negative, green links are positive.*

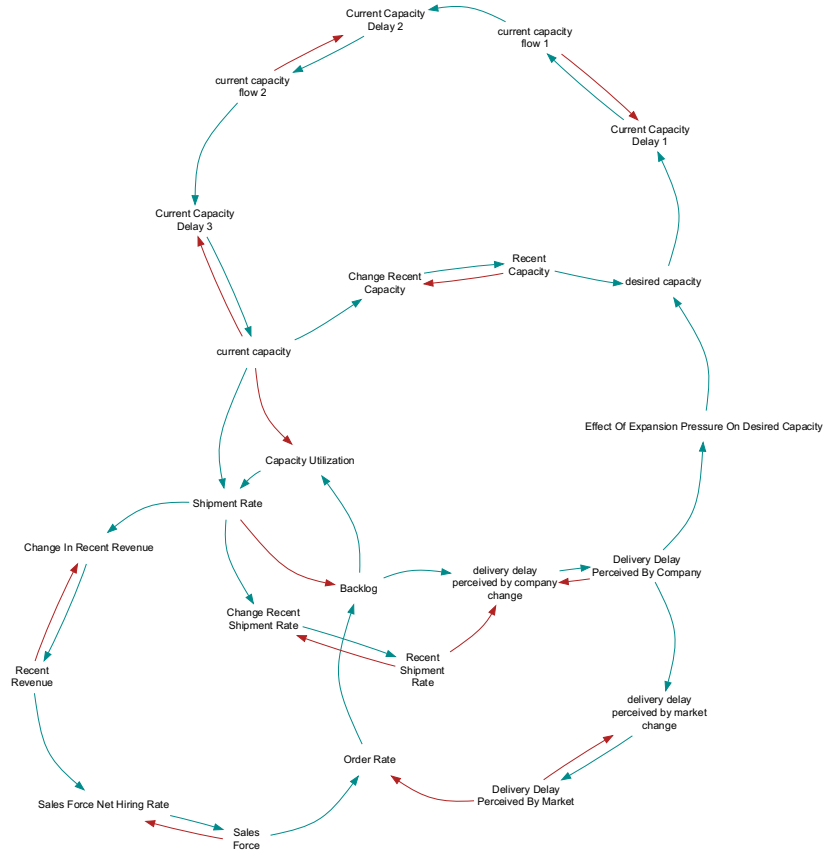

*Figure 6: Autogenerated Simplified CLD of Forrester's 1968 market growth model (link inclusion threshold 100%, loop inclusion threshold 0%.) Red links are negative, green links are positive.*

The full CLD in Figure 5 is large and difficult to understand for those who are not familiar with Forrester's model. Figure 6 is much simpler and in all ways is superior to the full CLD. It contains less than half of the variables of Figure 5, but still portrays the full feedback complexity of the model. Figure 6 was generated using a link inclusion threshold of 100%. All stocks and flows in the model were kept in Figure 6 because the loop inclusion threshold was set to 0%. This diagram is still complex because it represents all 23 feedback loops and contains 10 stocks, 12 flows and 2 auxiliaries. To further simplify the diagram, stocks and flows need to be removed, which will reduce the feedback complexity of the simplified CLD.

Setting the loop inclusion threshold to 20% while keeping the link inclusion threshold at 100% generates the diagram seen in Figure 7 which contains 7 stocks, 10 flows, and 2 auxiliaries. The maximally simplified CLD is pictured in Figure 8 where the both the link and loop threshold are set to 100%. In its most simplified form, the simplified CLD contains 4 variables, 2 flows 2 auxiliaries. The major tradeoff across Figure 5 and Figure 8 is the loss of descriptive power vs. the ease of cognition. This decision is best made on an individual by individual basis based on specific goals.

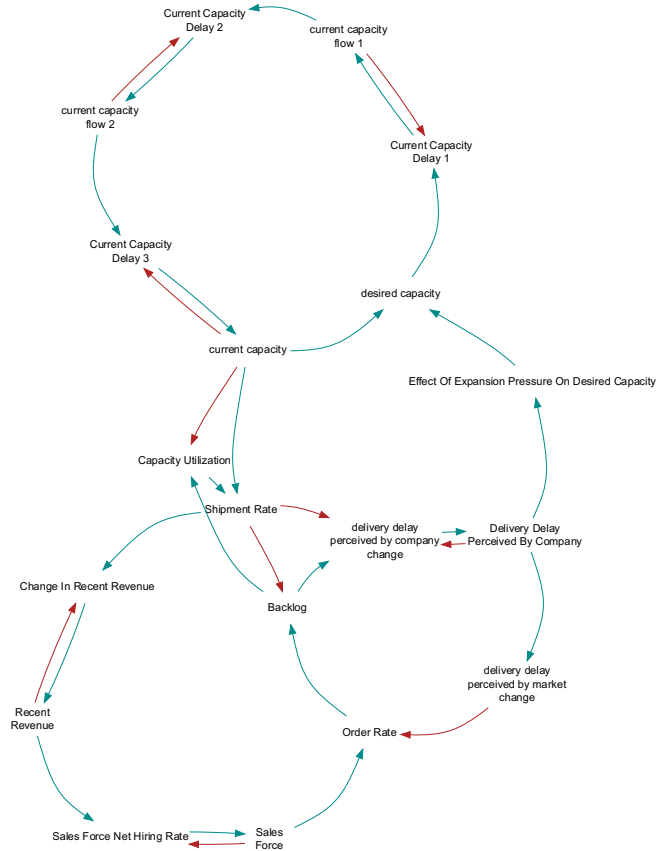

*Figure 7: Autogenerated Simplified CLD of Forrester's 1968 market growth model (link inclusion threshold 100%, loop inclusion threshold 20%.) Red links are negative, green links are positive.*

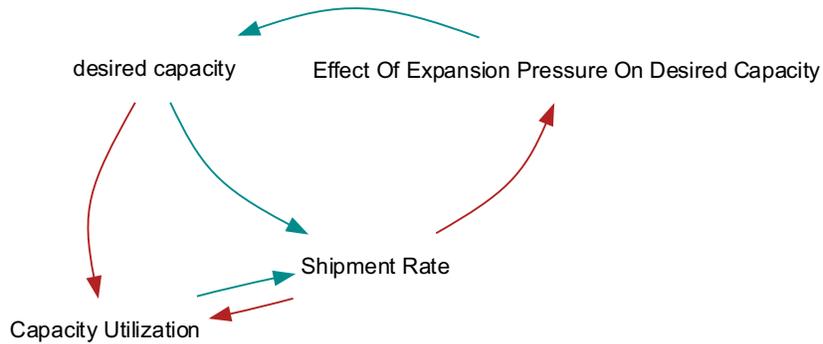

*Figure 8: Autogenerated Simplified CLD of Forrester's 1968 market growth model (link inclusion threshold 100%, loop inclusion threshold 100%). Red links are negative, green links are positive.*

As just demonstrated, when the simplification parameters are increased, fewer variables and therefore fewer feedback loops are presented to the end user. The link inclusion threshold typically removes complexity in the form of excess auxiliaries, and the loop inclusion threshold removes stocks, flows and feedback loops. These parameters provide the end user with tools to reshape the feedback loop complexity of the model on demand. Coupled with the machine generation of CLDs end users can now find CLDs which match their cognitive abilities and can be sure that the presented CLDs capture the most relevant portions of the feedback complexity of the model for that desired level of complexity.

3. *How can the results of an LTM analysis be easily visualized and communicated?*

By incorporating the results on an LTM analysis with an SFD or simplified CLD users can get information about the importance of feedback loops and links directly in the context of structure in an easy to interpret and understand way. By making SFDs and simplified CLDs animate, it becomes possible to visualize the feedback loop dominance profile of a model as the behavior unfolds in a way which is more directly related to the structure of the model as opposed to a comparative line graph.

In the animated SFDs or CLDs produced by LoopX, regardless of the type of diagram, color is used to represent the polarity of any link and thickness is used to represent the magnitude of the relative link score. In simplified CLDs where the links are built from causal pathways the relative link scores are multiplied to generate a relative link score for the simplified link. Users can change the diagrammatic representation of the model at any point in time via a simple dropdown box to allow for the most intuitive diagram to be presented. Users are able to obtain plots of behavior overtime for any variable in any diagram, as well as relative link scores and relative loop scores for any element by just clicking on the variable or connector or loop identifier in the diagram. All generated data is also offered for download in CSV form for external analysis and plotting.

On all diagrams, a table of relative loop scores is plotted showing the instantaneous contribution of each loop as well as by the relative average magnitude of the contribution over the entire model run of each loop to the behavior of all stocks in the model. This table is sorted by the relative average magnitude of the contribution to make the most important loops rise to the top. The loop identifier for any loop may be pressed, and, while held, highlights all variables and links in any of the diagrams that the loop represents, - removing all dominance information while doing so. This allows users to quickly identify what the meaning of all of the identified loops are and track them through any rendering of model structure as their placement does change since each diagram generation is a totally independent process as of the current writing of this paper.

An animation timeline is provided so that users may scrub through the visualization of loop dominance and pin it at any point in time to examine the state of the model (structurally or behaviorally) at that specific dt. Users are also free to adjust the link and loop inclusion thresholds at any point during the simulation as results are animating, or while they're scrubbing through results to explore the various levels of complexity in the explanations of model behavior. Figure 9 and Figure 10 depict the animation in both stock and flow and CLD diagrams of the bass diffusion model.

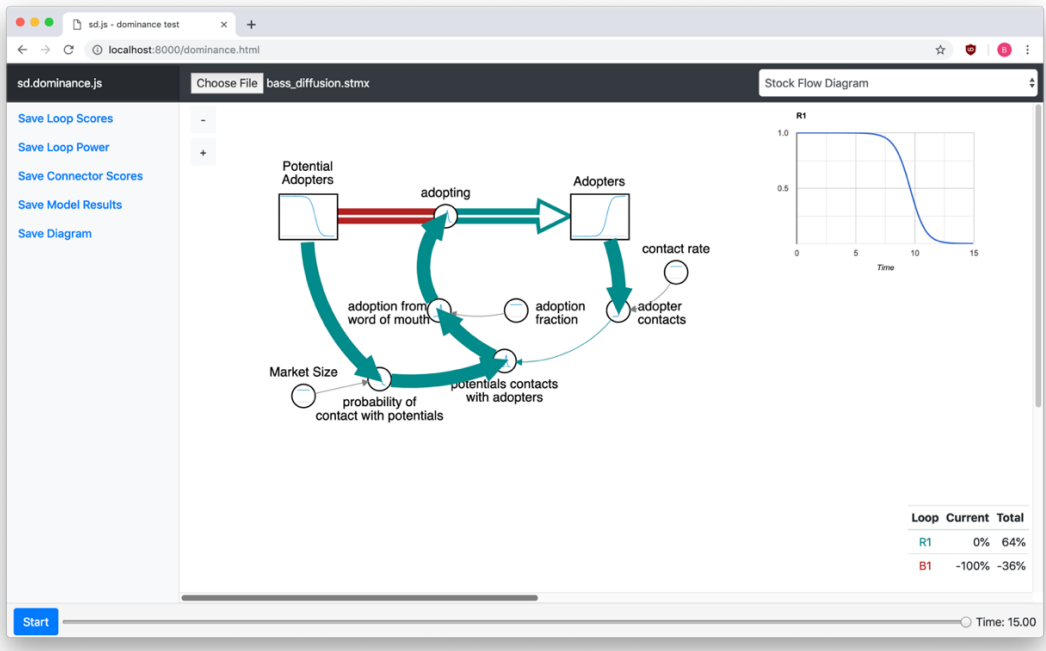

*Figure 9: Screenshot of LoopX showing a stock and flow diagram of the Bass Diffusion model at the final time period. Notice the coloring of the split flow 'adopting'. Red links are negative, green links are positive.*

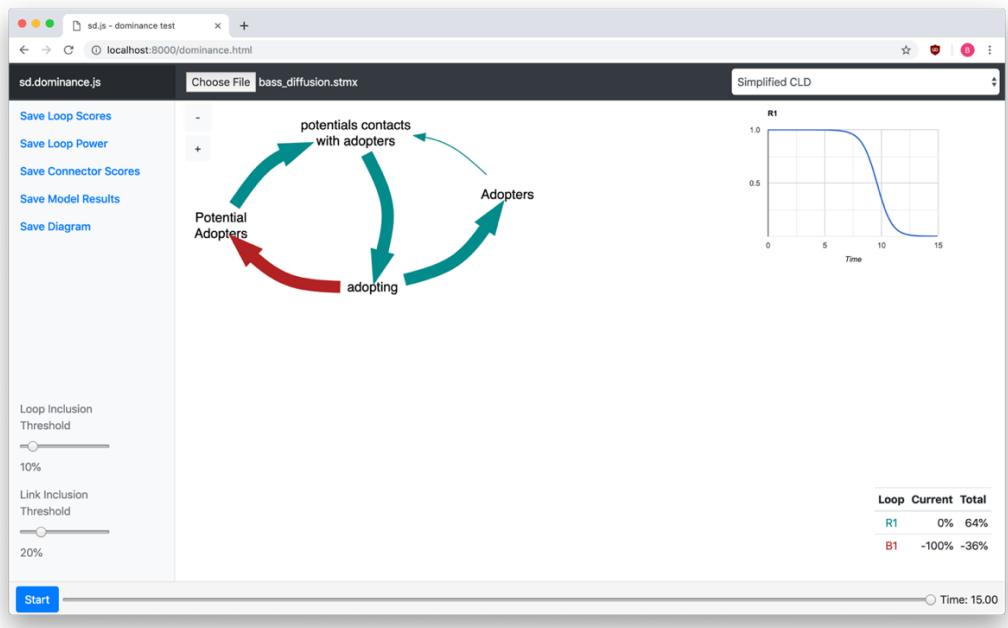

*Figure 10: Screenshot of LoopX showing a simplified CLD of the Bass Diffusion model at the final time period. Notice how Potential Adopters is driving potentials contacts with adopters signifying the importance of B1. Red links are negative, green links are positive.*

An interesting challenge in visualizing link importance and polarity in SFDs, is the case of flows. Flows are often times connected to two stocks and therefore have an in-polarity and contribution and an out-polarity and contribution.  In these cases (bass diffusion model), it would be non-sensical to render the whole flow with a single polarity and contribution because it is nearly always guaranteed that the flow has opposite polarities and often times it has different contributions.  To solve this problem, it makes the most amount of sense to split the flow in half, rendering the pipe sections before and after the flow valve with the polarity and contribution associated with the connecting stock.  This produces flows which make it very clear to users of the SFD of the hidden information links they contain. This is especially true re. the outflow from stocks where there is no arrowhead from the flow to the stock.  This can be observed in Figure 9.

**Discussion**

CLD generation with neato and the improved edge curving algorithm is successful because the generated CLDs appear to be naturally drawn, yet they tend to follow best practices as laid out by Richardson (1986).  Also, the generated CLDs tend to minimize the instances of non feedback looking like feedback while keeping short loops close in 2D space.  This enables users of these diagrams to enjoy not only all of the benefits of curved edged diagrams as measured by Xu et. al, 2012, but in addition, feedback becomes much easier to identify at a glance.  Finally, the techniques used to produce machine generated CLDs described in this paper can be applied independently from LTM and any other techniques discussed in this paper to generate high quality CLDs from network data.

The process for generating simplified CLDs works well because it maintains consistency of information regardless of aggregation level.  Because each aggregated link is composed of a specific and known list of disaggregated links, animation and visualization of the relative link score is not affected, because the link score is designed to be multiplied.  Because the aggregated diagrams are fully accurate representations of the relationships between the variables selected, there is confidence that information loss is minimal.  Regardless all information loss is controlled directly by the user via the choice of value for the loop inclusion threshold parameter.  The loop inclusion threshold will allow for information loss when it is set such that the stocks of specified unimportant feedback loops are removed from the diagram.  This means that the representation of those feedback loops in the simplified CLD can be lost if those stocks are not also resident in feedback loops of more importance.

The link inclusion threshold rarely leads to the loss of feedback loops in the aggregated diagrams because of the tendency of systems to generate large fluctuations in link score when approaching and leaving equilibrium states.  Therefore, the link inclusion threshold exceedingly rarely tends to be the source of removal for stocks and their associated feedback loops from the aggregated diagram.  The utility of the link inclusion threshold is to very quickly identify any relationships in the model that serve to expand the number of variables for reasons of equation simplification, - as opposed to for reasons of dynamic complexity.  It acts as a surgical scalpel for cutting away all of the variables in the model that do not serve as the interface between feedback loops, - allowing users to be presented with diagrams that contain a minimum number of variables representing a maximal amount of dynamic complexity.

Simplified CLDs are potentially useful in a wide variety of contexts.  The first is education where having access to accurate, but simpler depictions of structure could enhance learning opportunities.  The second is during model construction, where simplified CLDs could be used by model authors to verify their understanding of the most important dynamics driving behavior in their models.  Other use cases include presenting overviews of key model structures to policy makers.  Generally, simplified CLDs are useful in any situation where someone may want to explain a model, but not have to step through the full stock and flow diagram.

The simplified CLD process may also be useful for model simplification before running the ADAS algorithm (Schoenenberger et. al., 2017) to reduce model structure complexity before attempting to pattern match system archetypes.  The ADAS algorithm ought to do a better job of finding system architypes in their reduced forms.  This should be possible because the ADAS algorithm will not need to consider so many different possible mutations of structure.   Also, if the network searched, was limited by the loop and link inclusion thresholds then the results would also be limited to the structures which are provably most relevant to the model.  This would ensure that the algorithm presents the most comprehensive list of matches that are the most relevant to system behavior.

Simplified CLDs though are not without their problems.  Currently there is no indication of the quality of the generated diagram, and there is the potential for oversimplified explanations of behavior to be produced.  Currently the only way to combat this is by manually checking to see how many of the important feedback loops as determined by their average magnitude of the relative loop score, are actually resident in the generated CLD and evaluating the utility of the CLD based on that.  A second important problem is that the simplified links presented, while known to exist in the full model structure, may not be the most important causal pathway between the two variables.  This is because a simplified link may represent many different causal pathways, each with its own different importance and the simplification method only chooses the first one and presents that as if it were the only pathway.  This is a more significant problem to work around as it requires examining larger diagrams to specifically track the simplification as it happens.

The choice of the relative link magnitude normalized at each time step for each set of independent-to-dependent variables requires further explanation.  A normalized value is required for the animation of connectors and flows because a maximum thickness needs to be set. Otherwise thickness would have no bar to measure it against.  Without normalization, using the link score would create links whose thickness explodes towards infinity just as the link score does when models pass through or reach equilibrium states.

Another potential other choice for the source of data for the rendered link thickness could be an approach which would apply loop scores, sizing all of the links in a loop equally to better emphasize loop dominance.  The problem with animating the loop score arises often in models of any significance where important loops are derivatives of each other, sharing many links in common.   The trouble in these cases boils down to how to represent and display the information that a link is resident in multiple loops, - each with their own level of significance. Theorized techniques include drawing multiple links, one for each loop, or flashing through representations over time (per each time step) in proportion to loop contribution. Realistically though, any techniques chosen to represent loop score over links resident in multiple loops, will not scale with model complexity.  This problem is, moreover, compounded by the aggregation techniques presented. Because aggregate links, by their very nature, tend to be resident simultaneously in even more loops.

Other possible options for animating link thickness includes normalization of the link score to all link score values in the entire model at that specific time step, or normalization of the link score to all link score values in the entire model across all time steps.  The potential benefit of normalizing link score across all links at all times, is to offer an impression of the activity level of the model as a whole overtime.  This would make very clear when the model is reaching equilibrium, - as links would then tend to get thicker during these periods.  The problem, though, is that link scores even when plotted on a logarithmic scale appear to be exponential in shape during equilibrium events.  This means that dynamics would be all but impossible to observe at any time except for during equilibrium events because connectors and flows would constantly remain tiny.  Normalization across all links at a specific time period suffers from the same general issue, except its applied not to the diagram over time, but to parts of the diagram that reach equilibrium marginally slower or faster than other parts. That would give a very misleading perception of the model, exhibiting drastic shifts in dominance from one time step to the next, which is not borne out by the data.

The flaws in using relative link score magnitude as the source of the animation of connector thickness is that loops are not made any more easily identifiable and that the loop power is completely unobservable because values are normalized at each time step.  Ultimately, though, these downsides are mitigated via the loop legend that allows for quick and easy access to loop scores and, in the future, to loop power information over time, as well as for a quick and easy way to identify any loop of interest.

**Conclusions**
In the final chapter of Business Dynamics (2000), Sterman issued many challenges for the future of system dynamics, three of which have been answered by the creation of LoopX.  The first challenge; "Automated identification of dominant loops and feedback structure", has been answered previously by other techniques including EEA, PPM. But, for this first time, one of these automated loop dominance analysis techniques has been automated and packaged in such a way that the outcomes are easily accessible to a wide swath of practitioners in the field.  The second and third challenges; "Visualization of model behavior" and "Linking behavior to generative structure", also have a long past set of accomplishments.  LoopX represents a major success because it integrates loop dominance analysis techniques with model aggregation and visualization.  LoopX produces high quality, easy-to-decipher, animated SFDs and high-quality machine generated animated CLDs of the origins of model behavior via the integration of the results of an automated loop dominance analysis done by LTM.

At the current date, LoopX represents only a start to what ultimately may be possible. Efforts must be undertaken to measure the effectiveness of these techniques for teaching purposes, practitioner purposes and, potentially, after future revisions for use by the general public, before any definitive statements can be made about achieving Sterman's goals. Problems still need to be addressed include the scalability across giant models of a size such as T-21 or its brethren, which must include a significant re-engineering effort focused on deriving efficient solutions to the process of finding the simplified links. The ultimate viability of these techniques will be proven via their adoption in mainstream tooling.

**References:**
Chernobelskiy, R., Cunningham, K. I., Goodrich, M. T., Kobourov, S. G., & Trott, L. (2011, September). Force-directed Lombardi-style graph drawing. In *International Symposium on Graph Drawing* (pp. 320-331). Springer, Berlin, Heidelberg.

Eades, P. 1984. A heuristic for graph drawing. *Congressus numerantium*, *42*, 149-160.

Eberlein, RL. 1989. Simplification and understanding of models. *System Dynamics Review*, *5*(1), 51-68.

Forrester, JW. 1968. *Market growth as influenced by capital investment*. Industrial Management Review.

Gansner, ER. 2014. Using Graphviz as a Library (cgraph version). *published online August 21.*

Gansner, ER, & Hu, Y. 2010. Efficient, proximity-preserving node overlap removal. In *Journal of Graph Algorithms and Applications* (pp. 53-74).

Kamada, T., & Kawai, S. 1989. An algorithm for drawing general undirected graphs. *Information processing letters*, *31*(1), 7-15.

Kampmann CE. 2012. Feedback loop gains and system behaviour (1996). *System Dynamics Review 28*(4): 370–395.

North, S. C. 2004. NEATO user's guide. *Murray Hill, NJ: AT&T Bell Laboratories*.

Oliva, R. 2004. Model structure analysis through graph theory: partition heuristics and feedback structure decomposition. *System Dynamics Review*, *20*(4): 313-336.

Oliva R. 2016. Structural dominance analysis of large and stochastic models. *System Dynamics Review 32*(1): 26-51

Powers, R. (2019). SD.js in-browser system dynamics model simulation and display. GitHub repository: https://github.com/bpowers/sd.js

Richardson, G. P. (1986). Problems with causal-loop diagrams. *System dynamics review*, *2*(2), 158-170.

Saysel, AK, & Barlas, Y. (2006). Model simplification and validation with indirect structure validity tests. *System Dynamics Review*, *22*(3), 241-262.

Schoenberg, W. (2009). The Effectiveness of Force Directed Graphs vs. Causal Loop Diagrams: An experimental study. In *The 27th International Conference of the System Dynamics Society*.

Schoenberg, W, Davidsen, P, Eberlein, R, 2020. Understanding model behavior using loops that matter. *Under Review at the SDR*

Schoenenberger, Lukas, Alexander Schmid, and Markus Schwaninger. "Towards the algorithmic detection of archetypal structures in system dynamics." *System Dynamics Review* 31.1-2 (2015): 66-85.

Schoenenberger, L., Schmid, A., Ansah, J., & Schwaninger, M. 2017. The challenge of model complexity: improving the interpretation of large causal models through variety filters. *System Dynamics Review*, *33*(2), 112-137.

Sterman JD. 2000. *Business Dynamics: Systems Thinking and Modeling for a Complex World.* Irwin/McGraw-Hill, Boston.

Xu, K., Rooney, C., Passmore, P., Ham, DH., & Nguyen, PH. 2012. A user study on curved edges in graph visualization. *IEEE Transactions on Visualization and Computer Graphics*, *18*(12), 2449-2456.